\documentclass[preprint,superscriptaddress,showpacs,preprintnumbers]{revtex4-1}
\usepackage{amsfonts,amsmath,amssymb}
\usepackage{graphicx,color}
\usepackage{rotating}
\usepackage{dcolumn}
\usepackage{bm,ulem}
\usepackage[colorlinks=true, pdfstartview=FitV, linkcolor=blue, 
citecolor=blue,urlcolor=navyblue]{hyperref}

\begin{document}

\title{Constraints and correlations of nuclear matter parameters from a density-dependent van der 
Waals model}

\author{M. Dutra}
\affiliation{Departamento de F\'isica, Instituto Tecnol\'ogico de Aeron\'autica, DCTA, 
12228-900, S\~ao Jos\'e dos Campos, SP, Brazil}

\author{B. M. Santos}
\affiliation{Universidade Federal do Acre, 69920-900, Rio Branco, AC, Brazil}

\author{O. Louren\c{c}o}
\affiliation{Departamento de F\'isica, Instituto Tecnol\'ogico de Aeron\'autica, DCTA, 
12228-900, S\~ao Jos\'e dos Campos, SP, Brazil}

\date{\today}

\begin{abstract}
A recently proposed density-dependent van der Waals model, with only $4$ free parameters adjusted 
to fix binding energy, saturation density, symmetry energy, and incompressibility, is analyzed 
under symmetric and asymmetric nuclear matter constraints. In a previous paper, it was shown that 
this model is fully consistent with the constraints related to the binary neutron star merger event 
named GW170817 and reported by the LIGO and Virgo collaboration. Here, we show that it also 
describes satisfactorily the low and high-density regions of symmetric nuclear matter, with all the 
main constraints satisfied. We also found a linear correlation between the incompressibility and 
the skewness parameter, both at the saturation density and show how it relates to the crossing 
point presented in the incompressibility as a function of the density. In the asymmetric matter 
regime, other linear correlations are found, namely, the one between the symmetry energy ($J$) and 
its slope ($L_0$), and other one establishing the symmetry energy curvature as a function of the 
combination given by $3J-L_0$.
\end{abstract}

\maketitle

\section{Introduction}
\label{int}

An often approach used to treat many nucleons systems is to fit directly some of the many-nucleon 
observables, allowing the construction of thermodynamic equations of state to study the infinite 
nuclear matter. Among the main models constructed through this procedure, one can mention the widely 
known nonrelativistic Skyrme model~\cite{Skyrme1956,Skyrme1959}. For the relativistic case, on the 
other hand, a Lagrangian density is proposed and all thermodynamic quantities are derived from it. 
In its simplest version, the Walecka model \cite{wale74,sero79,walecka2}, based on relativistic 
field theory in a mean-field approach, depends on free parameters fitted to reproduce the infinite 
nuclear matter bulk properties. The applications of these models extend to different ranges of 
temperature and density. For the zero temperature regime, the detailed knowledge of the hadronic 
equations of the state, coming from both, relativistic and nonrelativistic models, is very important 
for the description of, for example, neutron stars, which are studied in densities up to around six 
times the nuclear saturation density ($\rho_0$)~\cite{stone-stev,glend}. 

In the finite temperature regime, in which the hadronic models are generalized to $T\ne 0$ but 
keeping the adjustment of the free parameters performed at $T = 0$, the phenomenon of phase 
transitions in nuclear matter takes place. In general, hadronic models exhibit a liquid-gas phase 
transition characterized by regions presenting low (gas phase) and high (liquid phase) densities at 
a temperature range of $T\lesssim 20$~MeV~\cite{critic1,critic2}. This is the typical feature 
presented by the known van der Waals (vdW) model~\cite{greiner,landau} in a temperature range of 
$T<T_c$, where $T_c$ is the critical temperature. Such similarities pose the question of whether the 
vdW model could also be used to describe hadronic systems. In Ref.~\cite{vov2}, the vdW model had 
its canonical ensemble thermodynamics converted into the grand canonical one. Later on, in 
Ref.~\cite{vov1}, the authors performed a direct application of this model to the nuclear matter 
environment at zero and finite temperature regime. 
%
%
In Ref.~\cite{vov3}, the authors used the same procedure adapting it to other real gases 
models (Redlich-Kwong-Soave, Peng-Robinson, and Clausius) to describe the 
infinite nuclear matter. However, these models present a limitation at high-density regime since 
they produce equations of state in which causality is violated for densities around $2.5\rho_0$ at 
most. Therefore, an important nuclear matter constraint, namely, the flow 
constraint~\cite{danielewicz}, can not be reached since it is defined in the region of $2\leqslant 
\rho/\rho_0\leqslant 5$. Furthermore, the stellar matter can also not be described due to this 
limitation. The origin of such a problem is the absence of a suitable relativistic treatment of the
hard-core repulsion, namely, the implementation of the Lorentz contraction (see, for instance, 
Ref.~\cite{bugaev2008}).

In Ref.~\cite{apj}, the authors considered a density-dependent vdW model in which the attractive and 
repulsive strengths were converted from constants to density-dependent functions properly chosen in 
order to avoid superluminal equations of state at low densities. It was shown that such a model, 
named as the \mbox{DD-vdW} model, is able to reproduce the flow constraint and also observational 
data of neutron stars. In particular, it was also shown that the constraints coming from the 
analysis of the GW170817 event, performed by the LIGO and Virgo collaboration, are fully satisfied 
by this model~\cite{ligo17,ligo18}. In the present work, we focus on the analysis of the 
\mbox{DD-vdW} model against the symmetric and asymmetric nuclear matter constraints used in 
Ref.~\cite{rmf} to test $263$ parametrizations of different relativistic mean-field (RMF) models. We 
also investigate the correlations between the bulk parameters presented by this model concerning the 
isoscalar and isovector sectors. 

In Sec.~\ref{real}, we present the density-dependent perspective of the real gases models with 
the \mbox{DD-vdW} model properly described in Sec.~\ref{ddvdw}. The model is submitted to the 
constraints in Sec.~\ref{constcorr}, and the summary and conclusions of this work are shown in 
Sec.~\ref{summ}.

\section{Real gases in a density dependent model perspective (nuclear matter description)}
\label{real}

The classical vdW model in the canonical ensemble is expressed by the following equation of 
state for the pressure~\cite{greiner,landau},
\begin{eqnarray}
P(\rho,T) = \frac{T\rho}{1-b\rho} - a\rho^2,
\label{vdwclass}
\end{eqnarray}
in which the parameters $a$ and $b$ represent, respectively, the strength of the attractive and 
repulsive parts of the interaction between the hard-sphere particles of radius $r$. The excluded 
volume parameter $b$ relates to $r$ through $b = 16\pi r^3/3$, in the so-called exclude volume 
mechanism. The Redlich-Kwong-Soave~\cite{rks1,rks2}, the Peng-Robinson~\cite{pr} and the Clausius 
models present the same expression for the first term on the right hand side of 
Eq.~(\ref{vdwclass}), but different structures for the second one. 

In Refs.~\cite{vov1,vov3}, a suitable conversion of Eq.~(\ref{vdwclass}) to the grand canonical 
ensemble, with relativistic treatment, was performed in order to use the vdW model and the other 
real gases ones in the description of infinite nuclear matter. By taking such a procedure into 
account, it is possible to construct a unique formulation to all the real gases if one considers a 
suitable description in terms of density-dependent functions for the attractive and repulsive parts 
of the nuclear interaction. In this case, the energy density for the infinite symmetric nuclear 
matter (SNM) in the grand canonical ensemble reads~\cite{apj}
\begin{eqnarray}
\mathcal{E}(\rho)=[1-b(\rho)\rho]\mathcal{E}^*_{\rm id}(\rho^*) - a(\rho)\rho^2,
\label{dedd}
\end{eqnarray}
with
\begin{eqnarray} 
\rho^* = \frac{\rho}{1-b(\rho)\rho}.
\label{rhocs}
\end{eqnarray}
The quantity $\mathcal{E}^*_{\rm id}$ is the kinetic energy of a relativistic ideal Fermi gas of 
nucleons of mass $M=938$ MeV, given by
\begin{eqnarray}
\mathcal{E}^*_{\rm id}(\rho^*) = 
\frac{\gamma}{2\pi^2}\int_0^{k_F^*}dk\,k^2(k^2+M^2)^{1/2},
\label{eidsym}
\end{eqnarray}
with $k_F^*=(6\pi^2\rho^*/\gamma)^{\frac{1}{3}}$. The degeneracy factor is $\gamma=4$ for SNM.

\begin{figure}[!htb]
\centering
\includegraphics[scale=0.4]{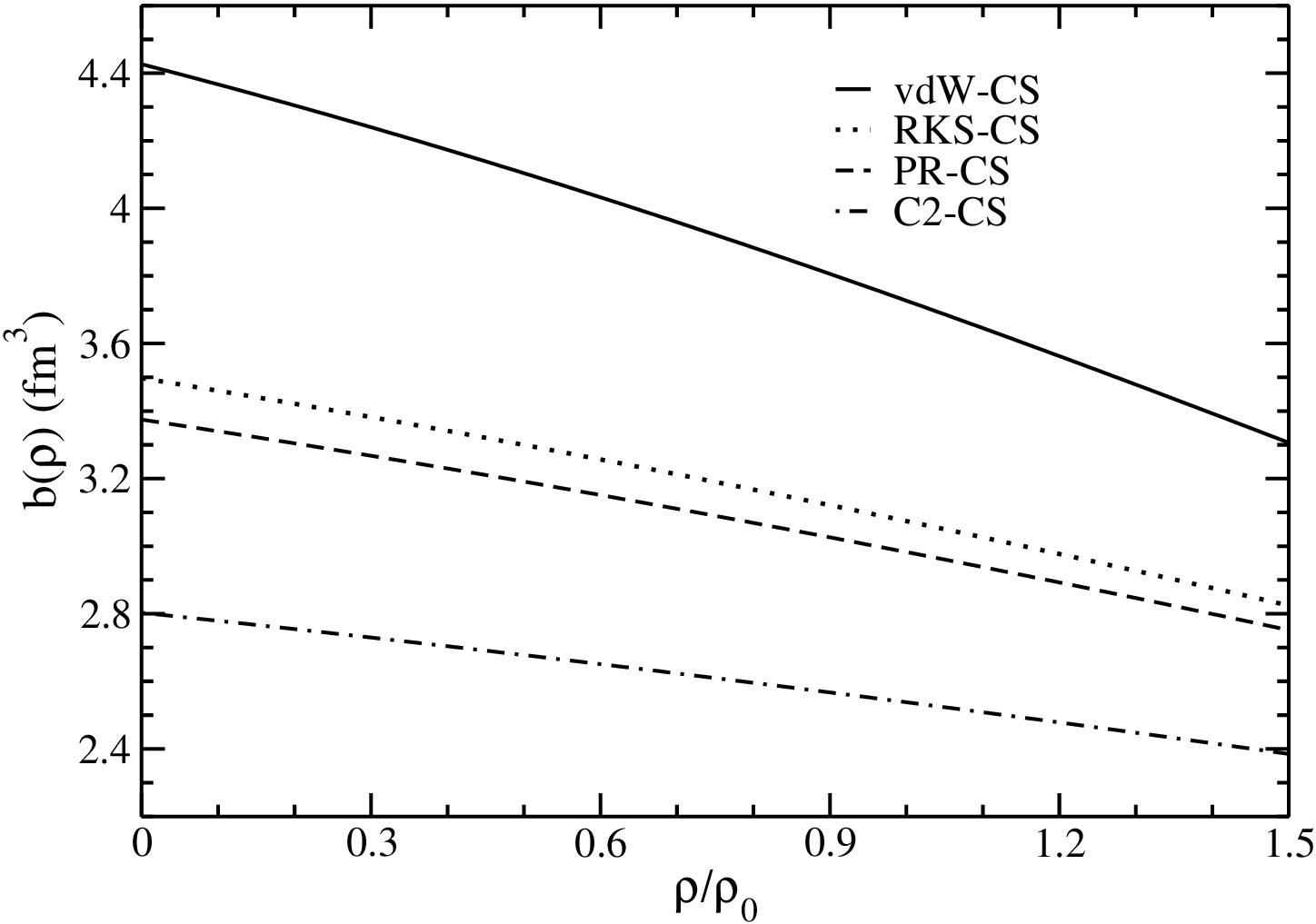}
\caption{Density dependence of the repulsive interaction for the real gases submitted to the 
CS method of excluded volume, Eq.~(\ref{brho}).}
\label{brhofig}
\end{figure}

The attractive interaction, now depending on $\rho$, assumes different forms for the real gases, 
namely,
\begin{eqnarray}
a(\rho)&=&a \hspace{5.35cm}{\rm (vdW)},
\label{arhovdw}\\
a(\rho)&=&\frac{a}{b\rho}{\rm ln}(1 + b\rho)\hspace{3.35cm} {\rm (RKS)},
\label{arhorks} \\
a(\rho)&=&\frac{a}{2\sqrt{2}b\rho}{\rm 
ln}\left[\frac{1+b\rho(1+\sqrt{2})}{1+b\rho(1-\sqrt{2})}\right]\hspace{0.65cm}
{\rm (PR)},
\label{arhopr}\\
a(\rho)&=&\frac{a}{1+b\rho}\hspace{4.45cm} {\rm (C2)},
\label{arhoc2}
\end{eqnarray}
for van der Waals (vdW), Redlich-Kwong-Soave (RKS), Peng-Robinson (PR), and Clausius-2 (C2) models. 
The last model is denoted by Clausius-2 since it is a two parameters version of the original 
Clausius model in which three parameters are considered~\cite{vov4}. For the repulsive interaction, 
on the other hand, it is possible to construct at least two possible forms for all the real gases 
models. The first is related to the conventional excluded volume mechanism in which 
\begin{eqnarray}
b(\rho)=b,
\label{bconst}
\end{eqnarray}
i. e., a pure constant. Another form takes into account the Carnahan-Starling (CS) \cite{cs} method 
of excluded volume, in which the pressure of hard-core nucleons of radius $r$ is given as $P=\rho T 
Z_{\rm CS} (\eta)$, with  
\begin{eqnarray} 
Z_{\rm CS} (\eta) = \frac{1 + \eta + \eta^2 - \eta^3}{(1-\eta)^3} 
= 1 + \sum_{j=0}^\infty(j^2+3j)\eta^j,
\end{eqnarray}
and $\eta=b\rho/4$. Using this method, we can find the first eight coefficients of the virial 
expansion unlike the traditional excluded volume method (EV), in which only two of them are 
recovered since, for this case, one has $Z(\eta)=(1-4\eta)^{-1}$. Such a procedure, used for the 
description of 
nuclear matter by the real gases models in Ref.~\cite{vov3}, leads to the following 
density-dependent form for the repulsive interaction~\cite{apj},
\begin{eqnarray}
b(\rho) = \frac{1}{\rho}-\frac{1}{\rho}{\rm exp} 
\left[\dfrac{-(4-\frac{3b\rho}{4})\frac{3b\rho}{4}}{\left(1-\frac{3b\rho}{4}\right)^2}
\right].
\label{brho}
\end{eqnarray}
As one can see in Fig.~\ref{brhofig}, $b(\rho)$ is a decreasing function for all the real gases 
models submitted to the CS procedure.

For all the real gases models constructed with the EV or CS methods, there are only two free 
parameters to be adjusted, namely, the constants $a$ and $b$, found in this case by imposing the 
binding energy value of $B_0 \approx 16.0$~MeV at the saturation density given by $\rho_0 \approx 
0.16$ fm$^{-3}$. For the vdW-EV model, for instance, one has \mbox{$a = 328.93$~MeV.fm$^3$} and $b 
= 3.41$~fm${^3}$ whereas for the vdW-CS model, these numbers change to $347.02$~MeV.fm$^3$ and 
$4.43$~fm${^3}$, respectively.

From Eq.~(\ref{dedd}), and by using the general relationship
\begin{eqnarray}
\frac{\partial}{\partial\rho} = 
\frac{1+b'\rho^2}{[1-b(\rho)\rho]^2}\frac{\partial}{\partial\rho^*},
\end{eqnarray}
coming from Eq.~(\ref{rhocs}), the pressure of the system is obtained, namely, 
\begin{eqnarray}
P(\rho) &=& \rho^2\frac{\partial(\mathcal{E}/\rho)}{\partial\rho}
= \rho^2\frac{\partial}{\partial\rho}[\mathcal{E}^*_{\rm id}/\rho^* - 
a(\rho)\rho]
=[1-b(\rho)\rho]^2\rho^{*2}\frac{\partial(\mathcal{E}^*_{\rm 
id}/\rho^*)}{\partial\rho^*}
- a'\rho^3 - a(\rho)\rho^2\nonumber\\
&=&(1+b'\rho^2)P^*_{\rm id} - a'\rho^3 - a(\rho)\rho^2 
= P^*_{\rm id} - a(\rho)\rho^2 + \rho\Sigma(\rho),
\label{pressdd}
\end{eqnarray}
with
\begin{eqnarray}
P^*_{\rm id} = \frac{\gamma}{6\pi^2}\int_0^{k_F^*}\frac{dk\,k^4}{(k^2+M^2)^{1/2}}.
\label{pidsym}
\end{eqnarray}
Here, $a'$ and $b'$ are the density derivatives of $a(\rho)$ and $b(\rho)$, respectively, and 
the rearrangement term is
\begin{eqnarray}
\Sigma(\rho) = b'\rho P^*_{\rm id} - a'\rho^2.
\label{reasym}
\end{eqnarray}

A generalization of the equation of state given in Eq.~(\ref{dedd}) to asymmetric nuclear matter, 
i.e., a system in which $y\equiv \rho_p/\rho \neq 1/2$ ($\rho_p$ is the proton density), was 
proposed in Ref.~\cite{apj}. It consists of adding a term proportional to the squared difference 
between protons and neutrons densities, namely, $\rho_3 = \rho_p -\rho_n$, as used widely in some 
RMF models~\cite{rmf}. The individual components (nucleons) are also distinguished by their 
respective kinetic energies. The final form for the energy density in this perspective is given by
\begin{align}
\mathcal{E}(\rho,y)&=[1-b(\rho)\rho]\mathcal{E}^*_{\rm id}(\rho^*_p,\rho^*_n) - 
a(\rho)\rho^2 + d\rho_3^2 \nonumber\\
& =[1-b(\rho)\rho]\mathcal{E}^*_{\rm id}(\rho^*_p,\rho^*_n) 
 - a(\rho)\rho^2 + d(2y - 1)^2\rho^2,
\label{deddy}
\end{align}
where $\mathcal{E}^*_{\rm id}(\rho^*_p,\rho^*_n) = \mathcal{E}^{*p}_{\rm id}(\rho^*_p) 
+ \mathcal{E}^{*n}_{\rm id}(\rho^*_n)$, for $\mathcal{E}^{*i}_{\rm id}(\rho^*_i)$ 
following the same form as in Eq.~(\ref{eidsym}) with $\gamma=2$, $k_F^*\rightarrow 
k_F^{*i}$ and $\rho^*\rightarrow \rho^*_i$ ($i=p,n$). The different densities are related to each 
other by
\begin{eqnarray} 
\rho^*_p = \frac{\rho_p}{1-b(\rho)\rho},\qquad \rho^*_n = \frac{\rho_n}{1-b(\rho)\rho},
\end{eqnarray}
with $b(\rho)$ given in Eq.~(\ref{brho}). An interpretation for this new term in Eq.~(\ref{deddy}) 
is that it mimics the $\rho$ meson exchange between the finite structure nucleons.

\section{DD-vdW model}
\label{ddvdw}

An important limitation of the real gases models presented in the previous section is the 
production of superluminal equations of state at densities not so high in comparison with the 
saturation density. The maximum densities attained by these models immediately before the violation 
of the causal limit ($v_s^2=\partial P/\partial\mathcal{E}>1$) is presented in 
Table~\ref{tabmodels} for the EV and CS method of excluded volume.
\begin{table}[!htb]
\centering
\caption{Maximum density ratio, $\rho_{\mbox{\tiny max}}/\rho_0$, for the real gases models 
submitted to the two mechanisms of excluded volume, namely, EV and CS methods.}
\begin{ruledtabular}
\begin{tabular}{lcc}
Model & $\rho_{\mbox{\tiny max}}^{\mbox{\tiny EV}}/\rho_0$ & $\rho_{\mbox{\tiny 
max}}^{\mbox{\tiny CS}}/\rho_0$
\\
\hline
vdW & 1.38 & 1.69 \\
RKS & 1.58 & 2.07 \\
PR  & 1.65 & 2.16 \\
C2  & 1.84 & 2.51 \\
\end{tabular}
\label{tabmodels}
\end{ruledtabular}
\end{table}

It is worth to note that the repulsive interaction plays an important role in the causal limit of 
the models since its density-dependent version induces the violation of causality at higher 
densities. The physical reason of this finding is that the CS method weakens the repulsive 
interaction as a function of density, according to the results displayed in Fig.~\ref{brhofig}, 
producing results closer to those of an ideal gas of massive point-like nucleons. For this case, 
causality is not violated~\cite{apj}. Even with this effect, the CS method is still not able to 
generate equations of state for densities greater than around $2.5\rho_0$, which is the best case 
of the C2-CS model. This result does not allow an analysis of the nuclear matter at high-density 
regime. Motivated by this limitation, it was proposed in Ref.~\cite{apj} a new form for the 
attractive interaction, given by 
\begin{eqnarray}
a(\rho)&=&\frac{a}{(1+b\rho)^n}.
\label{arhodd}
\end{eqnarray}
It is inspired in the C2 model, where causality is violated at higher densities in comparison with 
the remaining models. The set of equations of state given in Eqs.~(\ref{dedd}) and~(\ref{pressdd}) 
for SNM, and (\ref{deddy}) for the asymmetric case, with the repulsive and attractive interactions 
given by Eqs.~(\ref{brho}) and~(\ref{arhodd}), respectively, was named as the \mbox{DD-vdW} 
model. Notice that for the particular cases of $n=0$, and $n=1$, the \mbox{vdW-CS} and \mbox{C2-CS} 
models are reproduced, respectively, see Eqs.~(\ref{arhovdw}) and~(\ref{arhoc2}). In 
Table~\ref{tabddvdw}, we show the limit density reached by this model for some values of the 
power~$n$.
\begin{table}[!htb]
\centering
\caption{Maximum density ratio, $\rho_{\mbox{\tiny max}}/\rho_0$, for some values of the $n$ 
power, Eq.~(\ref{arhodd}), of the DD-vdW model.}
\begin{tabular}{lc}
\hline\hline
$n$ & $\rho_{\mbox{\tiny max}}/\rho_0$ 
\\
\hline
0  & 1.69\\
1  & 2.51\\
2  & 3.74\\
3  & 5.16\\
4  & 6.61\\
\hline\hline
\end{tabular}
\label{tabddvdw}
\end{table}

The effect of the $n$ power in $a(\rho)$ is that it weakens the strength of the attractive 
interaction. Therefore, the model approaches the free Fermi gas of massive particles in which the 
condition $v_s^2<1$ is verified. The combined effect of the density-dependent parameters $a(\rho)$ 
and $b(\rho)$ enables the model to reach higher densities.

The additional free parameter of the \mbox{DD-vdW} model, $n$, is adjusted in order to correctly 
fix the value of the incompressibility at the saturation density, namely, $K_0=9(\partial 
P/\partial\rho)|_{\rho_0}$. For the real gases models in the CS method, this value is in the range 
of $333\mbox{ MeV}\leqslant K_0\leqslant 601\mbox{ MeV}$~\cite{vov3}. In 
Refs.~\cite{Sagun:2017eye,Bugaev:2018uum}, a formulation including induced surface tension was 
implemented in the van der Waals model. The resulting approach also satisfies the flow constraint 
and the maximum mass observational data for neutron stars.

\section{Constraints and correlations from the DD-vdW model}
\label{constcorr}

In this section, we proceed to analyze the DD-vdW model against the constraints used to select the 
$35$ parametrizations of different RMF models out of $263$ investigated in Ref.~\cite{rmf}. These 
selected parametrizations had their bulk and thermodynamical quantities compared to respective 
theoretical/experimental data from symmetric matter, pure neutron matter (PNM), and a mixture of 
both, namely, symmetry energy and its slope evaluated at the saturation density ($J$ and $L_0$), and 
the ratio of the symmetry energy at $\rho_0/2$ to its value at $\rho_0$. We also investigate 
whether correlations between the bulk parameters also arise in the isoscalar and isovector sectors 
of the DD-vdW model.

\subsection{Constraints in symmetric matter}

\begin{figure}[!htb]
\centering
\includegraphics[scale=0.4]{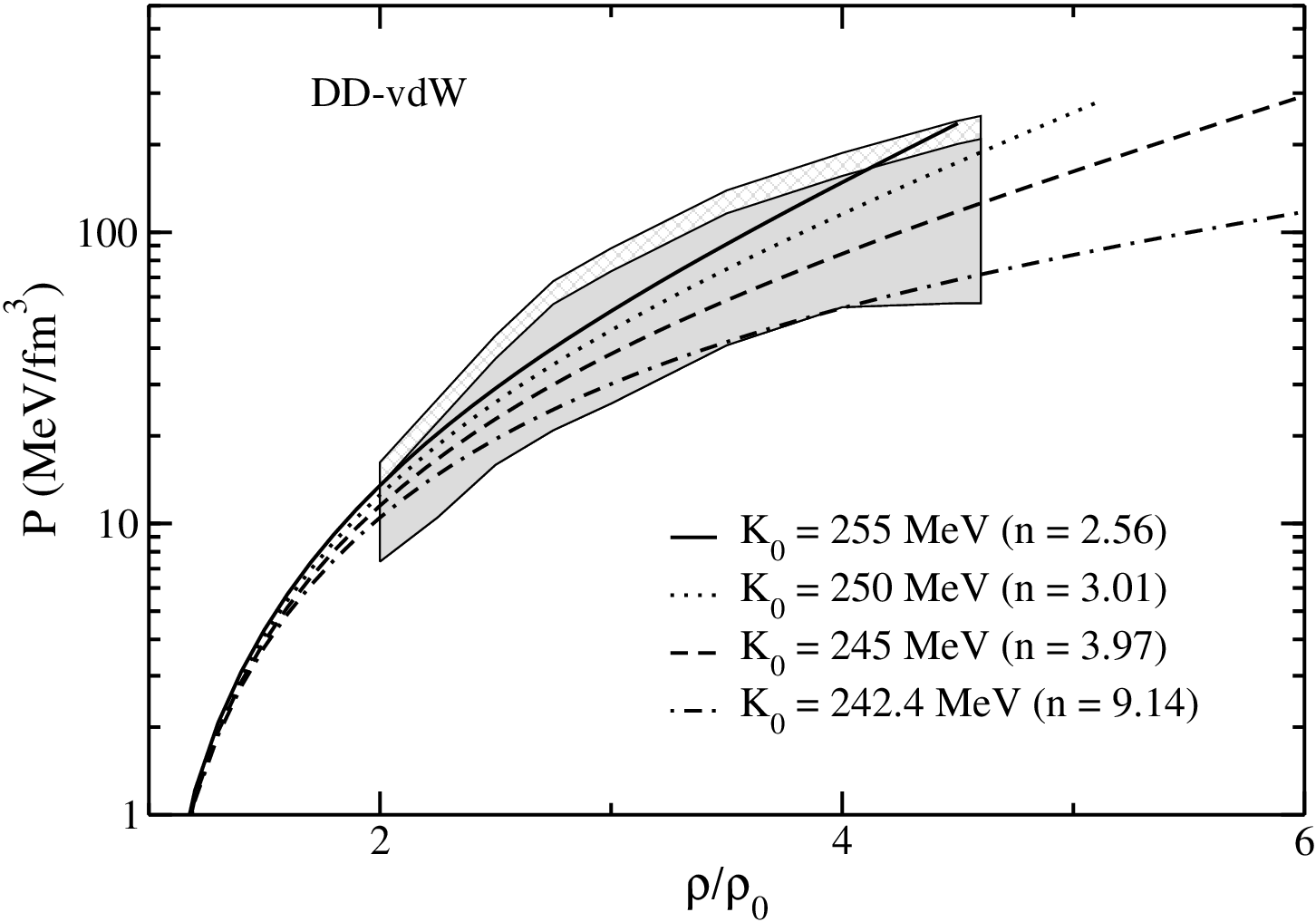}
\caption{Pressure as a function of $\rho/\rho_0$ for different \mbox{DD-vdW} parametrizations. 
Bands: flow constraint described in Refs.~\cite{rmf,danielewicz}.}
\label{flow}
\end{figure}

As a first constraint, we analyze the region in the density dependence of the pressure determined 
in Ref.~\cite{danielewicz} from the analysis of the flow in the collisions of \mbox{$^{197}\rm Au$}. 
The result is depicted in Fig.~\ref{flow}.

In Ref.~\cite{apj}, the authors considered the region obtained in Ref.~\cite{danielewicz}. Here, we 
also take into account the $20\%$ of increasing in this band as used in Ref.~\cite{rmf}. Such an 
increase was based on the band region obtained in Ref.~\cite{steiner} in which the authors performed 
an analysis based on observational data of bursting neutron stars showing photospheric radius 
expansion and transiently accreting neutron stars in quiescence. From the figure, one can verify 
that the parametrizations presenting $242.4\mbox{ MeV}\leqslant K_0 \leqslant 255\mbox{ MeV}$ are in 
full agreement with the flow constraint. Furthermore, such a range for $K_0$ also agrees with the 
constraint used in Ref.~\cite{rmf}, namely, $190\mbox{ MeV}\leqslant K_0 \leqslant 270\mbox{MeV}$. 
It also overlaps with the restriction for this quantity given by $250\mbox{ MeV}\leqslant K_0 
\leqslant 315\mbox{ MeV}$ found in Ref.~\cite{stone}. Furthermore, it is also consistent with the 
range recently discussed in Ref.~\cite{colo}, namely, $K_0=(240\pm 20)$ MeV.

\begin{figure}[!htb]
\centering
\includegraphics[scale=0.4]{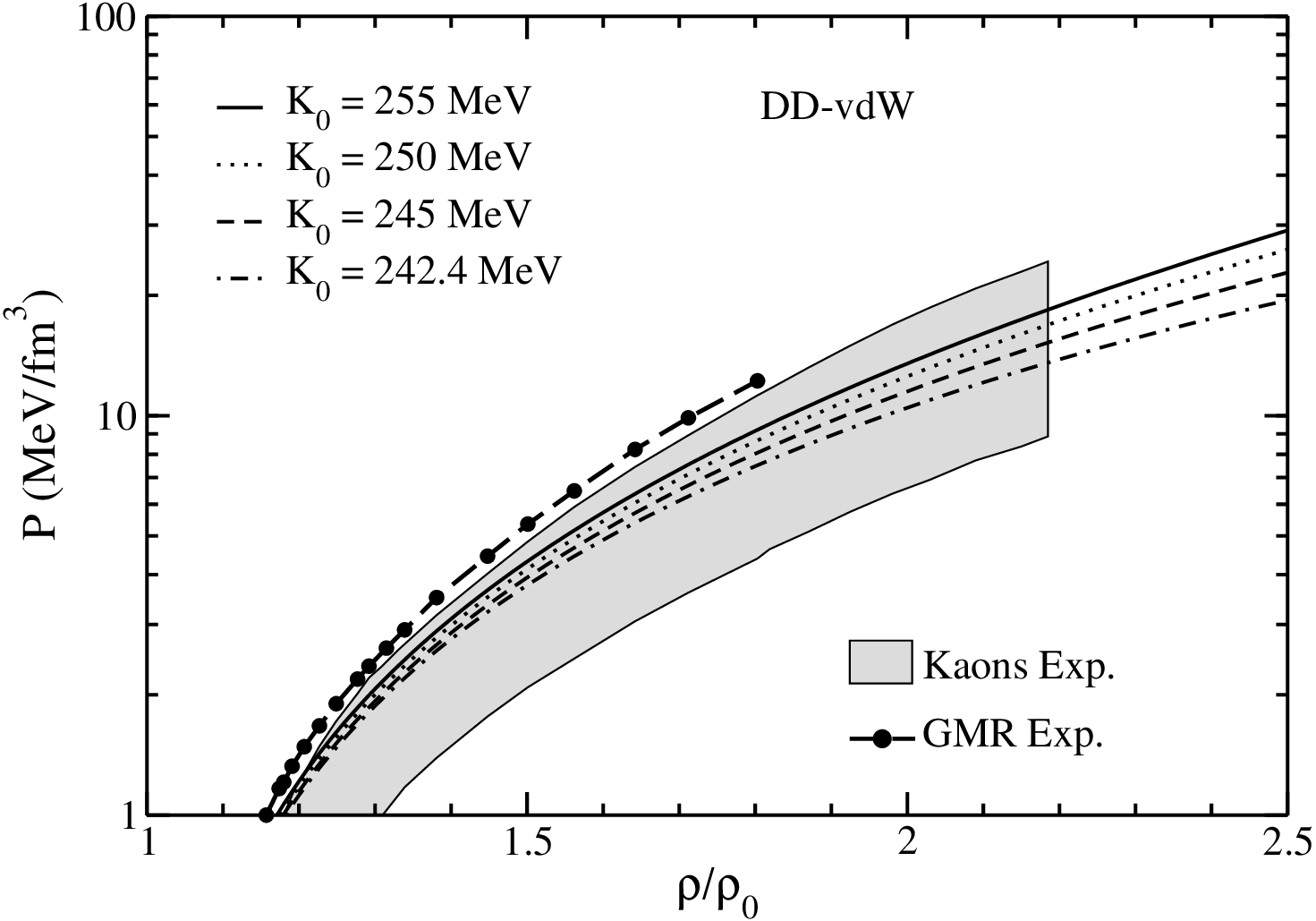}
\caption{Pressure as a function of $\rho/\rho_0$ for the \mbox{DD-vdW} parametrizations consistent 
with the flow constraint. Band and circle-dashed line: experimental data extracted from the 
Refs.~\cite{sm4a,sm4b}.}
\label{sm4}
\end{figure}

Another SNM constraint, coming from experiments of kaon production in heavy-ion collisions 
\cite{sm4a,sm4b}, also defines a band in the density dependence of the pressure but now at 
$1.2\leqslant\rho/\rho_0\leqslant 2.2$. In Fig.~\ref{sm4}, we show the behavior of the \mbox{DD-vdW} 
model against this constraint.

One can see that the \mbox{DD-vdW} model is also consistent with this particular constraint.

\subsection{Correlations in isoscalar sector}

We also investigate for the \mbox{DD-vdW} model, a possible correlation between bulk parameters in 
the symmetric nuclear matter, namely, the one involving $K_0$ and the skewness coefficient at the 
saturation density, $Q_0=Q(\rho_0)$ with
\begin{eqnarray}
Q(\rho)=27\rho^3\frac{\partial^3(\mathcal{E}/\rho)}{\partial\rho^3}.
\end{eqnarray}

The skewness coefficient is a bulk parameter that directly affects the high-density behavior of a 
hadronic model. In Ref.~\cite{skew}, for instance, the authors provide a detailed studied showing 
the impact of $Q_0$ in the RMF hadronic model. Different parametrizations with the same bulk 
parameters excepting the skewness coefficient were used to investigate the specific role played by 
$Q_0$ in the flow constraint and mass-radius diagram of neutron stars~\cite{skew}. 

\begin{figure}[!htb]
\centering
\includegraphics[scale=0.4]{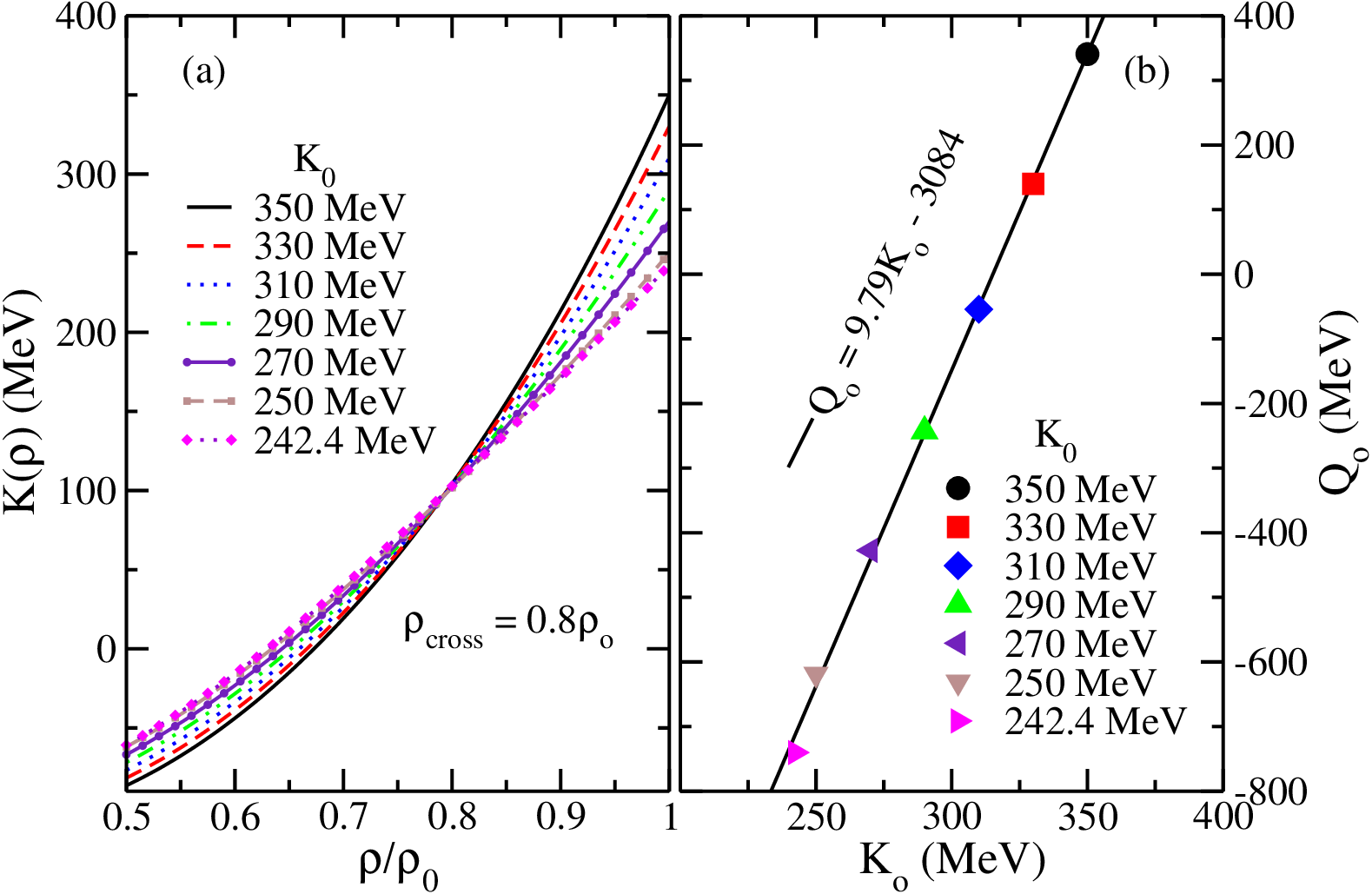}
\caption{(a) Incompressibility as a function of $\rho/\rho_0$ for some \mbox{DD-vdW} 
parametrizations. The density in which the parametrizations cross each other is given by 
$\rho_{\mbox{\tiny cross}}$. (b) $Q_0\times K_0$ linear correlation observed for the same 
parametrizations. 
Solid line: fitting curve.} 
\label{kcross}
\end{figure}

In order to examine a possible relationship between $Q_0$ and $K_0$, we proceed here as in 
Ref.~\cite{bianca2} where the authors have shown that a signature of linear correlations is 
exhibited in the density dependence of the bulk parameter analyzed. For instance, if we look for the 
density dependence of $K(\rho)$, and if a crossing point arises, consequently it generates to the 
linear relation between $Q_0$ and $K_0$. Actually, in Fig.~\ref{kcross} we show that this is the 
case for the \mbox{DD-vdW} model.

From the expansion of the energy per particle given by,
\begin{eqnarray}
\frac{\mathcal{E}}{\rho}\simeq -B_0  + \frac{K_0}{2!}x^2 + \frac{Q_0}{3!}x^3 + 
\cdots,
\end{eqnarray}
with $x=\frac{\rho-\rho_0}{3\rho_0}$, we found
\begin{align}
K(\rho) &= 18\rho\frac{\partial(\mathcal{E}/\rho)}{\partial\rho} + 
9\rho^2\frac{\partial^2(\mathcal{E}/\rho)}{\partial\rho^2} 
= 6(3x+1)\frac{\partial(\mathcal{E}/\rho)}{\partial x} + 
(3x+1)^2\frac{\partial^2(\mathcal{E}/\rho)}{\partial x^2} 
\nonumber\\
&\simeq(3x+1)\left[ K_0 + (9K_0+Q_0)x + 6Q_0x^2\right].
\end{align}
If the following linear correlation is true,
\begin{eqnarray}
Q_0 = a_1K_0 + a_2, 
\label{q0k0}
\end{eqnarray}
then
\begin{eqnarray}
K(\rho)\simeq (3x+1)[K_0F(x) + (1 + 6x)a_2x],
\end{eqnarray}
with $F(x)=1 + ( 9 + a_1 )x + 6a_1x^2$, and $K(\rho_{\mbox{\tiny cross}})$ will present the same 
value for different parametrizations only if one has $F(x_{\mbox{\tiny cross}})=0$ for a particular 
point (crossing point) given by $\rho_{\mbox{\tiny cross}} = (3x_{\mbox{\tiny cross}} + 1)\rho_0$. 
Since we found this crossing point in Fig.~\ref{kcross}{\color{blue}a}, the correlation in 
Eq.~(\ref{q0k0}) holds and the condition $F(x_{\mbox{\tiny cross}})=0$ is satisfied, in this case 
for $\rho_{\mbox{\tiny cross}}/\rho_0=0.8$ (also from Fig.~\ref{kcross}{\color{blue}a}). The 
relationship between $Q_0$ and $K_0$ is exhibited in Fig.~\ref{kcross}{\color{blue}b}, confirming 
the linear correlation. A linear fitting points out to $a_1=9.79$, generating the more accurate 
value of $\rho_{\mbox{\tiny cross}}/\rho_0=0.798$, coming from the exact solution of 
$F(x_{\mbox{\tiny cross}})=0$. The authors of Ref.~\cite{margueron} also found a crossing for some 
nonrelativistic Skyrme and Gogny parametrizations, with $\rho_{\mbox{\tiny cross}}/\rho_0=0.7$. 
However, for the RMF models analyzed, they did not find any linear correlations or even crossing 
points. In this case, this is due to the different values of the effective nucleon mass for the 
distinct parametrizations, as explained in Ref.~\cite{bianca2}. However, for Boguta-Bodmer models in 
which effective mass is $0.6M$, for instance, one has a crossing in the $K(\rho)$ curve at 
$\rho_{\mbox{\tiny cross}}/\rho_0=0.77$ and, as a consequence, a linear correlation between $Q_0$ 
and $K_0$ \cite{bianca2}. The crossing density found in the \mbox{DD-vdW} model is close to the 
values obtained by the aforementioned models.

For the \mbox{DD-vdW} parametrizations in which the flow constraint is satisfied, the respective 
$Q_0$ values are in the range of $-740~\mbox{MeV}\leqslant Q_0 \leqslant -569~\mbox{MeV}$, which is 
compatible with other calculations giving $-690~\mbox{MeV}\leqslant Q_0 \leqslant 
-208~\mbox{MeV}$~\cite{q01}, $-790~\mbox{MeV}\leqslant Q_0 \leqslant -330~\mbox{MeV}$~\cite{q01}, 
and $-1200~\mbox{MeV}\leqslant Q_0 \leqslant -200~\mbox{MeV}$~\cite{q02}.

\subsection{Asymmetric matter}
\label{asym}

In the symmetric nuclear matter, the \mbox{DD-vdW} model contains only three free parameters, 
namely, the constants $a$, $b$ and $n$ present in Eqs.~(\ref{brho}), and (\ref{arhodd}). In the 
generalization to asymmetric matter proposed in Eq.~(\ref{deddy}), there is one more free constant, 
namely, the $d$ parameter. In Ref.~\cite{apj}, this constant was adjusted in order to fix the 
symmetry energy at the saturation density, $J\equiv \mathcal{S}(\rho_0)$, where
\begin{align}
\mathcal{S}(\rho) =
\frac{1}{8}\frac{\partial^{2}(\mathcal{E}/\rho)}{\partial y^{2}}\Big|_{y=\frac{1}{2}} 
= \mathcal{S}_{kin}^*(\rho) + d\rho,
\label{esym}
\end{align}
with $\mathcal{S}_{kin}^*(\rho)=k_F^{*2}/(6E_F^*)$ and $E_F^* = \sqrt{k_F^{*2}+M^2}$. The choice of 
fixing $J$ in the range of $25~\mbox{MeV}\leqslant J \leqslant 35~\mbox{MeV}$ automatically becomes 
the model consistent with the constraint used in Ref.~\cite{rmf}, obtained from the data collection 
reported in Ref.~\cite{jl}.

From Eq.(\ref{esym}), one obtains the symmetry energy slope as follows,
\begin{equation}
L(\rho) = 3\rho\frac{\partial\mathcal{S}}{\partial\rho} 
= \xi(\rho)L_{kin}^*(\rho) + 3d\rho,
\end{equation}
where
\begin{align}
L_{kin}^*(\rho)=\frac{k_F^{*2}}{3E^*_F}\left(1 - \frac{k_F^{*2}}{2E_F^{*2}}\right) 
= 2\mathcal{S}_{kin}^*\left(1 - \frac{3\mathcal{S}_{kin}^*}{E_F^*}\right),
\end{align}
and $\xi(\rho) = (1 + b'\rho^2)/[1 - b(\rho)\rho]$.

One advantage of the specific form of the last term added in the energy density in 
Eq.~(\ref{deddy}), namely, that one containing the $d$ parameter, is that it generates an 
analytical relationship between $\mathcal{S}(\rho)$ and $L(\rho)$ for all densities, 
since one can write $d=(\mathcal{S}-\mathcal{S}_{kin}^*)/\rho$ from Eq.~(\ref{esym}). This 
result leads to $L(\rho) = 3\mathcal{S}(\rho) + g(\rho)$, with 
\begin{align}
g(\rho) &= \mathcal{S}_{kin}^*(\rho)\left\lbrace
2\xi(\rho)\left[ 1 - \frac{3\mathcal{S}_{kin}^*(\rho)}{E_F^*(\rho)}\right] - 3 
\right\rbrace
\end{align}
i. e., a linear correlation between $L(\rho)$ and $\mathcal{S}(\rho)$ is clearly established if 
$g(\rho)$ does not vary significantly. 

A possible linear correlation at saturation density given by
\begin{align}
L(\rho_0) \equiv L_0 = 3J + g_0
\label{jl}
\end{align}
is of great interest, since it is observed in many hadronic models~\cite{bianca2,baldo,bianca,lim}. 
It will be satisfied if $g(\rho_0)\equiv g_0$ is approximately fixed regarding a variation of $J$. 
Indeed, this is the case for the \mbox{DD-vdW} parametrizations in which $\rho_0$, $B_0$ and $K_0$ 
are kept fixed for different values of $J$. The reason is that $g_0$ is a function of quantities 
depending only on free parameters adjusted from observables related to the symmetric matter, which 
means that $g_0$ depends only on $\rho_0$, $B_0$ and $K_0$, i.e., $g_0=g_0(\rho_0,B_0,K_0)$. For 
parametrizations presenting these symmetric matter quantities fixed, independently of their $J$ 
values, $g_0$ is a constant. This situation could not be the same if we had proposed terms in 
Eq.~(\ref{deddy}) with more than one isovector free parameter to be adjusted. The linear correlation 
in Eq.~(\ref{jl}) can be blurred in this case since $g_0$ can also depend on $J$. However, some 
relativistic and nonrelativistic models with more than one isovector free parameter also present the 
same relationship of Eq.~(\ref{jl}), as the NL3*~\cite{nl3*} family and some Skyrme 
parametrizations, for instance~\cite{bianca2,fatt}.
\begin{figure}[!htb]
\centering
\includegraphics[scale=0.4]{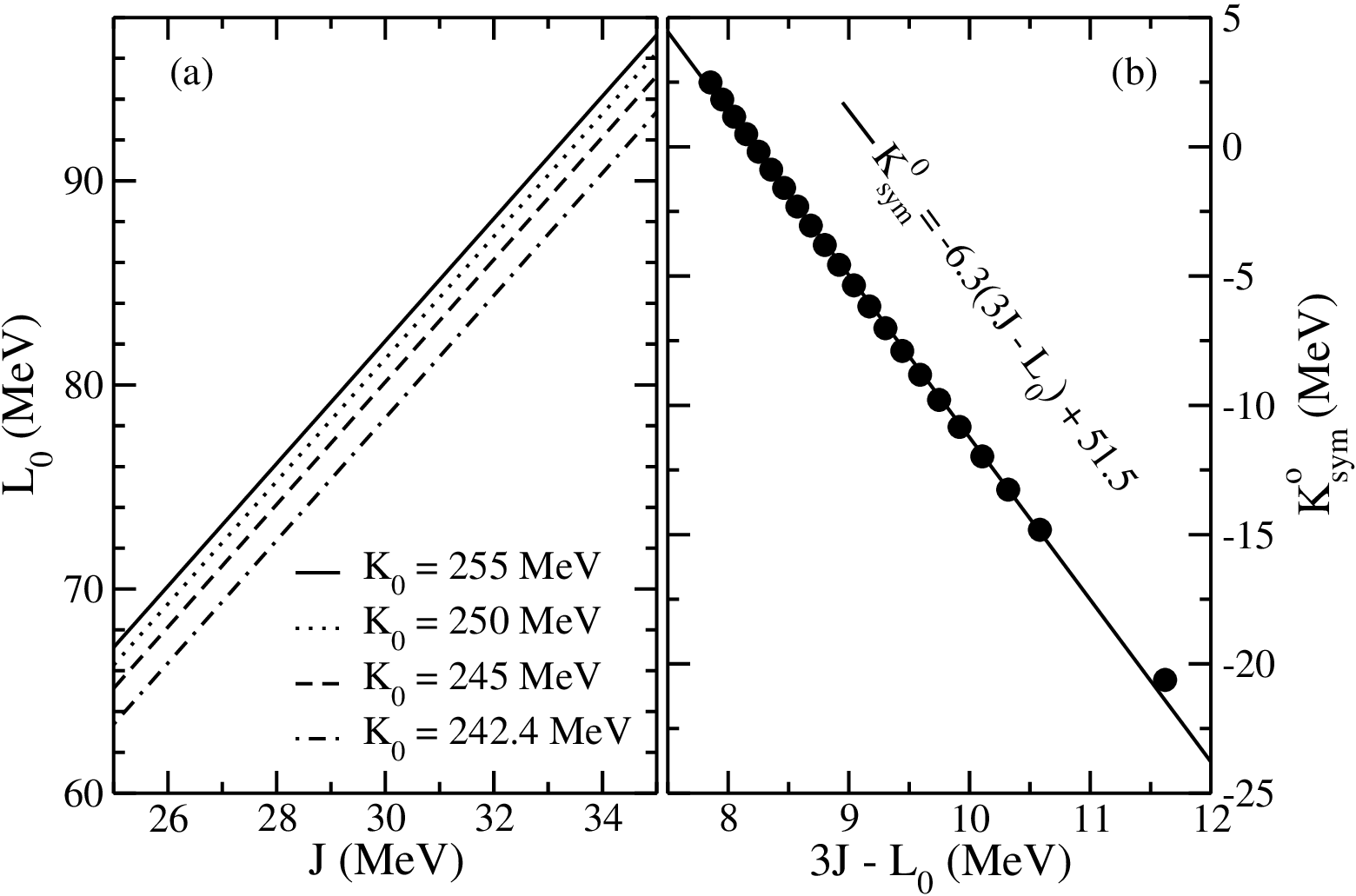}
\caption{(a) $L_0$ as a function of $J$, and (b) $K_{\mbox{\small sym}}^0$ as a function of 
$3J-L_0=-g_0(\rho_0,B_0,K_0)$. The full circles were obtained by using the range of $242.4\mbox{ 
MeV}\leqslant K_0 \leqslant 255\mbox{ MeV}$, with fixed values of $\rho_0=0.16$~fm$^{-3}$ and 
$B_0=16$~MeV. Both panels constructed from the \mbox{DD-vdW} parametrizations consistent with the 
flow constraint.}
\label{jlksym}
\end{figure}

For \mbox{DD-vdW} parametrizations consistent with the flow constraint, namely, those in which 
$242.4~\mbox{MeV}\leqslant K_0 \leqslant 255~\mbox{MeV}$, we found the slope of symmetry energy at 
saturation density in the range of $63.4~\mbox{MeV}\leqslant L_0 \leqslant 97.1~\mbox{MeV}$, see 
Fig.~\ref{jlksym}{\color{blue}a}. It was obtained from Eq.~(\ref{jl}) and by taking into account the 
$J$ constraint of $25~\mbox{MeV}\leqslant J \leqslant 35~\mbox{MeV}$. These $L_0$ values are in full 
agreement with constraint of $25~\mbox{MeV}\leqslant L_0 \leqslant 115~\mbox{MeV}$ applied in the 
RMF parametrizations studied in Ref.~\cite{rmf}, and obtained from the data collection presented in 
Ref.~\cite{jl}.

It is also possible to find analytic expressions for higher-order terms of $\mathcal{S}$ from 
Eq.~(\ref{esym}). In particular, its curvature is given by 
\begin{align}
K_{\mbox{\small sym}}(\rho) = 9\rho^2\frac{\partial^2\mathcal{S}}{\partial\rho^2}
= 3\rho g'(\rho).
\end{align}
At the saturation density, we have $K_{\mbox{\small sym}}^0 = 3\rho_0 g'_0$, with $K_{\mbox{\small 
sym}}^0\equiv K_{\mbox{\small sym}}(\rho_0)$. Likewise~$g_0$, the quantity $g'_0$ is a function 
only of the isoscalar bulk parameters of the model, i.e, $g'_0=g'_0(\rho_0,B_0,K_0)$. 

\begin{figure}[!htb]
\centering
\includegraphics[scale=0.4]{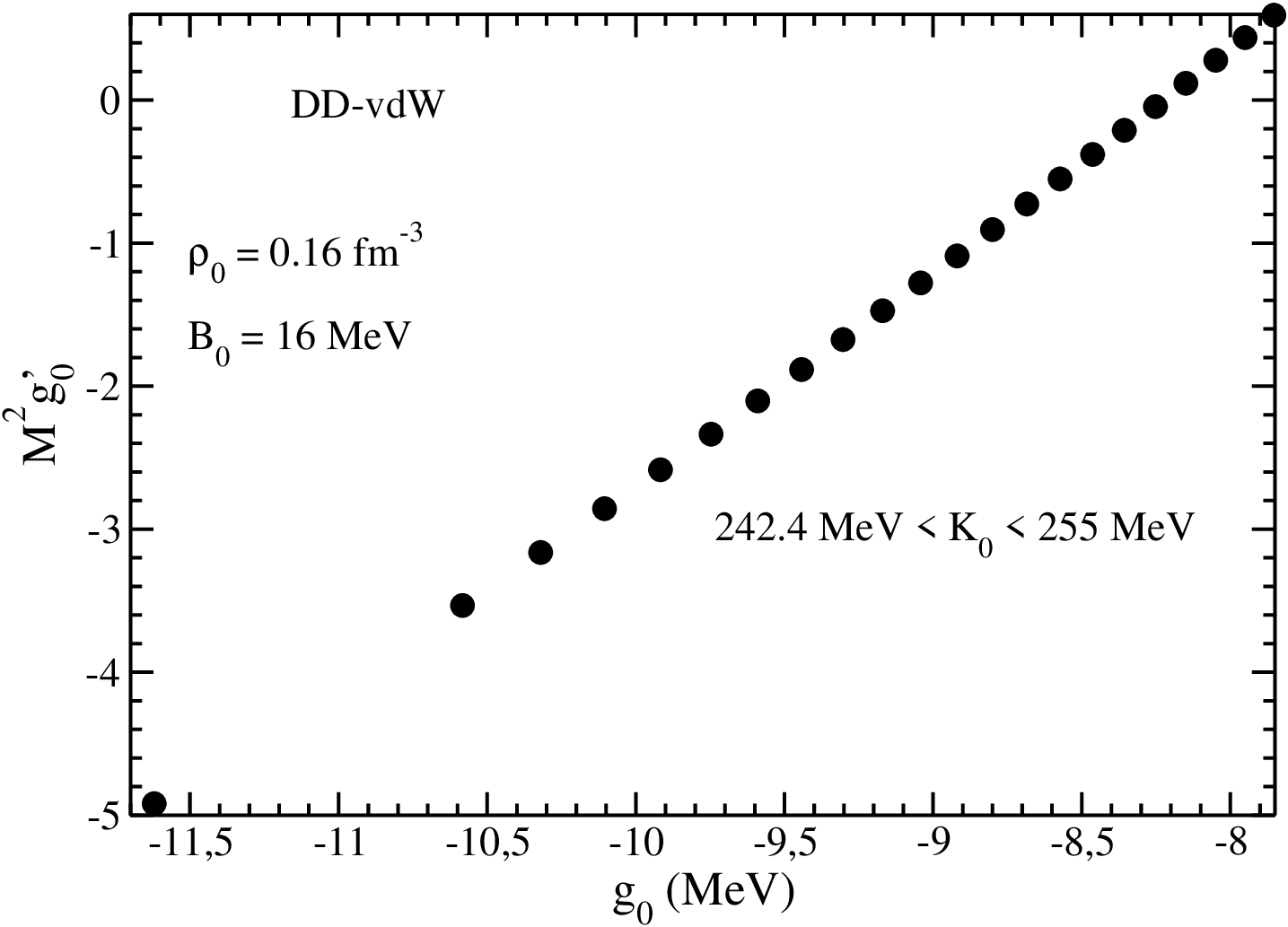}
\caption{Correlation between $M^2g'_0$ and $g_0$ for \mbox{DD-vdW} parametrizations 
consistent with the flow constraint.}
\label{glg}
\end{figure}

Notice that if the quantity $g'_0$ can be given also as a function of $g_0$ in a linear form 
as $g'_0 = \alpha_1g_0 + \alpha_2$, then $K_{\mbox{\small sym}}^0$ would be given by 
$K_{\mbox{\small sym}}^0=3\rho_0[-\alpha_1(3J - L_0)+\alpha_2]$, according to Eq.~(\ref{jl}), i.e., 
a linear correlation between $K_{\mbox{\small sym}}^0$ and $3J - L_0$ would arise 
($\alpha_1$~and~$\alpha_2$ are constants). Indeed, it is verified that $g'_0$ linearly depends on 
$g_0$ as one can see in Fig.~\ref{glg}. The black circles were obtained by using the values of 
$\rho_0=0.16$~fm$^{-3}$, $B_0=16$~MeV and running~$K_0$ in the range of $242.4\mbox{ MeV}\leqslant 
K_0 \leqslant 255\mbox{ MeV}$. 
The direct consequence of the relationship between $g'_0$ and $g_0$ is the correlation between 
$K_{\mbox{\small sym}}^0$ and $3J-L_0$ presented in Fig.~\ref{jlksym}{\color{blue}b}, with a 
fitting curve given by $K_{\mbox{\small sym}}^0 =  -6.3(3J-L_0) + 51.5$. As in Fig.~\ref{glg}, the 
full circles in Fig.~\ref{jlksym}{\color{blue}b} were calculated by using the range of $242.4\mbox{ 
MeV}\leqslant K_0 \leqslant 255\mbox{ MeV}$, with the aforementioned fixed values for $\rho_0$ and 
$B_0$. This linear correlation between $K_{\mbox{\small sym}}^0$ and $3J-L_0$ was 
shown for $500$ relativistic and nonrelativistic parametrizations in Ref.~\cite{mondal1}, and 
specifically for the Skyrme model in Ref.~\cite{mondal2}. Also in Ref.~\cite{margueron2}, the 
authors have discussed such relationships. From this strong linear behavior, it is possible to 
obtain a range for the symmetry energy curvature for the \mbox{DD-vdW} parametrizations consistent 
with flow constraint, namely, $-20.6~\mbox{MeV}\leqslant K_{\mbox{\small sym}}^0 \leqslant 
2.5~\mbox{MeV}$. This range is inside the one obtained for some RMF parametrizations in 
Ref.~\cite{mondal1}, according to Fig.~1 of this reference.

\begin{figure}[!htb]
\centering
\includegraphics[scale=0.4]{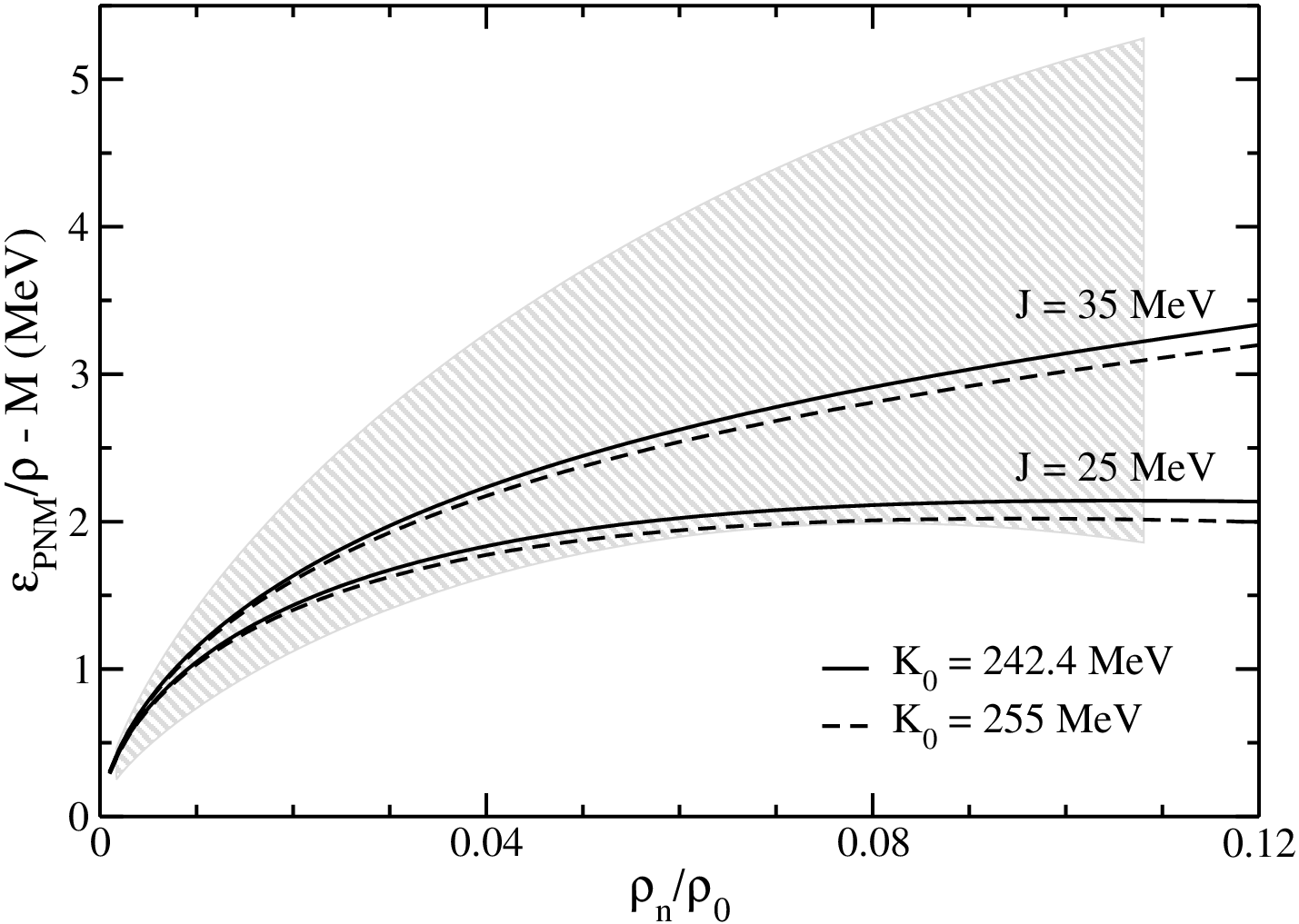}
\caption{Density dependence of energy per particle in PNM for \mbox{DD-vdW} 
parametrizations consistent with the flow constraint. Band region: constraint given in 
Refs.~\cite{rmf,skyrme}.}
\label{pnm1}
\end{figure}

Another constraint adopted in Refs.~\cite{rmf,skyrme} was the one concerning the density dependence 
of PNM energy per particle (Eq.~\ref{deddy} evaluated at $y=0$) at very low density 
regime. It was based on the lattice chiral effective theory including corrections due to finite 
scattering length, nonzero effective range, and higher-order corrections related to the 
nucleon-nucleon interaction in that regime. We submitted the \mbox{DD-vdW} model also to this 
constraint with results displayed in Fig.~\ref{pnm1}.
One can see that the \mbox{DD-vdW} model is also consistent with this particular restriction.

Finally, we analyze the \mbox{DD-vdW} model against a constraint explored in 
Refs.~\cite{rmf,skyrme} 
related to the difference between proton and neutron densities (neutron skin thickness). In 
Ref.~\cite{dani03}, the author associated this difference to the reduction of the symmetry energy at 
$\rho_0/2$ in terms of the quantity $r\equiv\mathcal{S}(\rho_0/2)/J$ and found the limit of 
$0.57\leqslant r \leqslant 0.83$. This constraint is named as MIX4 in Refs.~\cite{rmf,skyrme}. In 
Table~\ref{tabsj}, we show the values of $\mathcal{S}(\rho_0/2)/J$ for the parametrizations of the 
\mbox{DD-vdW} model consistent with the $J$ and the flow constraints.
\begin{table}[!htb]
\centering
\caption{Values of $r$ for \mbox{DD-vdW} parametrizations presenting $242.4\mbox{ MeV}\leqslant K_0 
\leqslant 255\mbox{ MeV}$ and $25\mbox{ MeV}\leqslant J \leqslant 35\mbox{ MeV}$.}
\begin{ruledtabular}
\begin{tabular}{ccc}
 $K_0$ (MeV) & $J$ (MeV) & $r$
\\
\hline
$242.4$  &  $25$  & $0.56$ \\
$242.4$  &  $35$  & $0.54$ \\
$255$    &  $25$  & $0.55$ \\
$255$    &  $35$  & $0.53$ \\
\end{tabular}
\label{tabsj}
\end{ruledtabular}
\end{table}

From Table~\ref{tabsj}, one can verify that all parametrizations present numbers out of the limit 
of the MIX4 constraint. However, to follow the same criterion adopted in Refs.~\cite{rmf,skyrme} for 
that parametrizations do not satisfy only one of the analyzed constraints, we also establish here 
that a particular parametrization is approved if the value of the quantity exceeds the limits of 
the respective constraint by less than $5\%$. This is the case for the parametrizations reported in 
Table~\ref{tabsj}, except for that presented in the last line. Therefore, one can conclude that the 
\mbox{DD-vdW} model also predicts parametrizations in agreement with this particular constraint.

\section{Summary and Conclusions}
\label{summ}

In this paper, we revisited the recently proposed density-dependent van der Waals model 
(\mbox{DD-vdW})~\cite{apj} by submitting it to the symmetric and asymmetric nuclear matter 
constraints used in Ref.~\cite{rmf}. The constraints used are summarized in Table~\ref{set2a} as 
follows.
\begin{table}[!htb]
\scriptsize
\caption{Set of updated constraints (SET2a) used in Ref.~\cite{rmf} and applied to the \mbox{DD-vdW} 
model. See that reference for more details concerning each constraint.}
\centering
\begin{ruledtabular}
\begin{tabular}{lccc}
Constraint & Quantity      & Density region       & Range of constraint \\ 
\hline
SM1    & $K_0$     & at $\rho_0$  & 190 $-$ 270 MeV \\
SM3a   & $P(\rho)$ & $2<\frac{\rho}{\rho_0}<5$ & Band Region \\
SM4    & $P(\rho)$ & $1.2<\frac{\rho}{\rho_0}<2.2$    & Band Region\\
PNM1   & $\mathcal{E}_{\mbox{\tiny PNM}}/\rho$ & $0.017<\frac{\rho}{\rho_{\rm
o}}<0.108$   & Band Region \\
MIX1a  & $J$       & at $\rho_0$  & 25 $-$ 35 MeV \\
MIX2a  & $L_0$     & at $\rho_0$  & 25 $-$ 115 MeV \\
MIX4   & $\frac{\mathcal{S}(\rho_0/2)}{J}$  & at $\rho_0$ and $\rho_0/2$ & 0.57 $-$ 0.83\\
\end{tabular}
\label{set2a}
\end{ruledtabular}
\end{table}

In order to become the model capable to reach the high density regime, the new density dependent 
attractive interaction is considered, see Eq.~(\ref{arhodd}). The new free parameter, $n$, is used 
to fix the value of the incompressibility at the saturation density. The other ones, found at the 
symmetric nuclear matter regime, namely, $a$ and $b$, are found by imposing the model to present the 
$B_0=16$~MeV at $\rho=\rho_0$. By imposing the model to satisfy the flow 
constraint~\cite{danielewicz}, see Fig.~\ref{flow}, $K_0$ is found to be restricted to $242.4\mbox{ 
MeV}\leqslant K_0 \leqslant 255\mbox{ MeV}$, values compatible with the SM1 constraint. It also 
overlaps with the restriction proposed in Ref.~\cite{stone}, namely, $250\mbox{ MeV}\leqslant K_0 
\leqslant 315\mbox{ MeV}$. 

The last free parameter, $d$, is adjusted to generate parametrizations in which $25\mbox{ 
MeV}\leqslant J \leqslant 35\mbox{ MeV}$. Therefore, the model automatically satisfies the 
constraint named as MIX1a. All the other constraints are fully satisfied, excepting the MIX4. 
However, some parametrizations are out of its range only by less than $5\%$, which makes this 
constraint also satisfied, according to the criterion adopted in Refs.~\cite{rmf,skyrme}.

We also analyzed the correlations between the bulk parameters in this model. It was shown that the 
crossing point in the $K(\rho)$ function, see Fig.~\ref{kcross}{\color{blue}a}, is related to the 
linear correlation between $K_0$ and the skewness coefficient at $\rho=\rho_0$, see 
Fig.~\ref{kcross}{\color{blue}b}. For the isovector sector, we also verified the linear correlation 
between $J$ and $L_0$ (slope parameter at $\rho=\rho_0$), according to Eq.~(\ref{jl}) and 
Fig.~\ref{jlksym}{\color{blue}a}. Furthermore, a linear correlation between $3J-L_0$ and 
$K_{\mbox{\small sym}}^0$ (symmetry energy curvature at $\rho=\rho_0$) also arises for the 
\mbox{DD-vdW} model, as shown in Fig.~\ref{jlksym}{\color{blue}b}. This kind of correlation was also 
studied for relativistic and nonrelativistic mean-field models in 
Refs.~\cite{mondal1,mondal2,margueron2}.

In Ref.~\cite{apj}, it was shown that this new proposed model, with only $4$ free parameters, 
satisfactorily describes the constraints related to the binary neutron star merger event named as 
GW170817, and reported by the LIGO and Virgo collaboration~\cite{ligo17,ligo18}. Here, we continued 
the analysis of the model and found that the mainly symmetric and asymmetric nuclear matter 
constraints are also satisfied, with some correlations between bulk parameters also observed. 

\section*{Acknowledgments} 

This work is a part of the project INCT-FNA Proc. No. 464898/2014-5, partially supported by 
Conselho Nacional de Desenvolvimento Cient\'ifico e Tecnol\'ogico (CNPq) under grants 310242/2017-7 
and 406958/2018-1 (O. L.) and 433369/2018-3 (M. D.), and by Funda\c{c}\~ao de Amparo \`a Pesquisa do 
Estado de S\~ao Paulo (FAPESP) under thematic projects 2013/26258-4 (O. L.) and 2017/05660-0 
(O. L., M. D.).

\end{document}